\documentclass[11pt,fleqn]{iopart}
\usepackage[utf8]{inputenc}
\usepackage[T1]{fontenc}
\usepackage{fouriernc}
\expandafter\let\csname equation*\endcsname=\relax
\expandafter\let\csname endequation*\endcsname=\relax
\usepackage{amsmath,amssymb,amsfonts,amsthm}
\usepackage{graphicx}
\numberwithin{equation}{section}
\begin{document}
\title[Superintegrable discrete oscillator and bivariate Meixner polynomials]{A superintegrable discrete oscillator and two-variable Meixner polynomials}
\author{Julien Gaboriaud}
\ead{julien.gaboriaud@umontreal.ca}
\address{D\'epartement de physique, Universit\'e de Montr\'eal, Montr\'eal (QC), Canada}
\author{Vincent X. Genest}
\ead{vincent.genest@umontreal.ca}
\address{Centre de recherches math\'ematiques, Universit\'e de Montr\'eal, Montr\'eal (QC), Canada}
\author{Jessica Lemieux}
\ead{jlemi049@uottawa.ca}
\address{D\'epartement de physique, Universit\'e d'Ottawa, Ottawa (ON), Canada}
\author{Luc Vinet}
\ead{luc.vinet@umontreal.ca}
\address{Centre de recherches math\'ematiques, Universit\'e de Montr\'eal, Montr\'eal (QC), Canada}
\begin{abstract}
A superintegrable, discrete model of the quantum isotropic oscillator in two-dimensions is introduced. The system is defined on the regular, infinite-dimensional $\mathbb{N}\times \mathbb{N}$ lattice. It is governed by a Hamiltonian expressed as a seven-point difference operator involving three parameters. The exact solutions of the model are given in terms of the two-variable Meixner polynomials orthogonal with respect to the negative trinomial distribution. The constants of motion of the system are constructed using the raising and lowering operators for these polynomials. They are shown to generate an $\mathfrak{su}(2)$ invariance algebra. The two-variable Meixner polynomials are seen to support  irreducible representations of this algebra. In the continuum limit, where the lattice constant tends to zero, the standard isotropic quantum oscillator in two dimensions is recovered. The limit process from the two-variable Meixner polynomials to a product of two Hermite polynomials is carried out by involving the bivariate Charlier polynomials.
\end{abstract}
\newpage
\section{Introduction}
The purpose of this paper is to present a discrete model of the two-dimensional quantum oscillator that is both superintegrable and exactly solvable. The wavefunctions of this system will be given in terms of the two-variable Meixner polynomials and the constants of motion will be seen to satisfy the $\mathfrak{su}(2)$ algebra.

A considerable amount of literature can be found on superintegrable systems and there is a sustained interest in enlarging the documented set of models with that property. Recall that a quantum system with $d$ degrees of freedom governed by a Hamiltonian $H$ is said to be super\-integrable if it possesses, including $H$ itself, $2d-1$ algebraically independent constants of motion, that is operators that commute with the Hamiltonian. The quintessential example of a quantum super\-integrable system is the two-dimensional harmonic oscillator, whose constants of motion generate the $\mathfrak{su}(2)$ algebra. One of the motivating observations behind the study of superintegrable systems is that they are exactly solvable, which makes them prime candidates for modeling purposes. Also of importance is the fact that these systems form a bedrock for the analysis of symmetries, of the associated algebraic structures and their representations, and of special functions. The majority of quantum superintegrable models cataloged so far comprises continuous systems, but there has also been some progress in the study of discrete systems \cite{2004_Levi&Tempesta&Winternitz_JMathPhys_45_4077,2013_Miki&Post&Vinet&Zhedanov_JPhysA_46_125207}; for a review on superintegrable systems (mostly continuous ones) and their applications, see \cite{2013_Miller&Post&Winternitz_JPhysA_46_423001}.

In the past years, several discrete models of the one-dimensional quantum oscillator, either finite or infinite, were introduced \cite{1998_Atakishiyev&Jafarov&Nagiyev&Wolf_RevMexFis_44_235, 2005_Atakishiyev&Pogosyan&Wolf_PhysPartNucl_36_247,  2012_Jafarov&VanderJeugt_JPhysA_45_485201, 2011_Jafarov&Stoilova&VanderJeugt_JPhysA_44_265203}. The most studied system, originally proposed in \cite{1997_Atakishiyev&Wolf_JOptSocAmA_14_1467} as a model of multimodal waveguides with a finite number of sensor points, has $\mathfrak{su}(2)$ as its dynamical algebra. In this model, the Hamiltonian, the position and momentum operators are expressed in terms of $\mathfrak{su}(2)$ generators, the eigenstates of the system are the basis vectors of unitary irreducible representations of $\mathfrak{su}(2)$ and the wavefunctions are expressed in terms of the one-variable Krawtchouk polynomials. As a result, the Hamiltonian has a finite number of eigenvalues and the spectra of the position and momentum operators are both discrete and finite. Germane to the present paper is also the discrete oscillator model based on the univariate Meixner polynomials and related to the $\mathfrak{su}(1,1)$ algebra considered in \cite{1998_Atakishiyev&Jafarov&Nagiyev&Wolf_RevMexFis_44_235}. See also \cite{2006_Klimyk_UkrJPhys_51_1019}, where instead the Meixner-Pollaczek polynomials are involved.

The Krawtchouk one-dimensional finite/discrete oscillator has been exploited to construct finite/discrete systems in two dimensions. Two approaches have been used. The first approach consist in taking the direct product of two one-dimensional $\mathfrak{su}(2)$ systems to obtain a system defined on a square grid with $\mathfrak{su}(2)\oplus \mathfrak{su}(2)$ as its dynamical algebra \cite{2001_Atakishiyev&Pogosyan&Vicent&Wolf_JPhysA_34_9381}. In the second approach \cite{2001_Atakishiyev&Pogosyan&Vicent&Wolf_JPhysA_34_9399}, the isomorphism $\mathfrak{so}(4) \cong \mathfrak{su}(2)\oplus \mathfrak{su}(2)$ is exploited to obtain a description of the finite oscillator on the square grid in terms of discrete radial and angular coordinates. In the continuum limit, both of these models tend to the standard two-dimensional oscillator. However, they do not exhibit the $\mathfrak{su}(2)$ invariance, or symmetry algebra, of the standard two-dimensional oscillator.

Recently  another discrete and finite model of the  two-dimensional oscillator was proposed  \cite{2013_Miki&Post&Vinet&Zhedanov_JPhysA_46_125207}. This model is defined on a triangular lattice of a given size and, like the standard oscillator in two dimensions, it is superintegrable and has $\mathfrak{su}(2)$ for symmetry algebra. The wavefunctions of the model, which support irreducible representations of $\mathfrak{su}(2)$ at fixed energy, are given in terms of the two-variable Krawtchouk polynomials introduced by Griffiths \cite{1971_Griffiths_AusJStat_13_27}. These polynomials of two discrete variables are orthogonal with respect to the trinomial distribution.  As required, this model tends to the standard two-dimensional oscillator in the continuum limit.  

Shortly after \cite{2013_Miki&Post&Vinet&Zhedanov_JPhysA_46_125207} appeared, it was recognized that the $d$-variable Krawtchouk polynomials of Griffiths arise as matrix elements of the unitary representations of the rotation group $SO(d+1)$ on oscillator states  \cite{2013_Genest&Vinet&Zhedanov_JPhysA_46_505203}. This interpretation has provided a cogent framework for the characterization of these orthogonal functions and has led to a number of new identities. The group theoretical interpretation was also extended to two other families of discrete multivariate polynomials: the multivariate Meixner and Charlier polynomials, orthogonal with respect to the negative multinomial and multivariate Poisson distributions, respectively. The multi-variable Meixner polynomials, also introduced by Griffiths \cite{1975_Griffiths_JMultVarAnal_5_271}, were shown to arise as matrix elements of unitary representations of the pseudo-rotation group $SO(d,1)$  on oscillator states \cite{2014_Genest&Miki&Vinet&Zhedanov_JPhysA_47_045207}. As for the multivariate Charlier polynomials, they were first introduced as matrix elements of unitary representations of the Euclidean group on oscillator states \cite{2014_Genest&Miki&Vinet&Zhedanov_JPhysA_47_215204}. Let us note that these family of multivariate polynomials also arise in probability theory in connection with the so-called Lancaster distributions \cite{Griffiths-2014}.
 
In this paper, we present a new discrete oscillator model in two-dimensions based on the two-variable Meixner polynomials. The model is defined on the regular infinite-dimensional $\mathbb{N}\times \mathbb{N}$ lattice. It is governed by a Hamiltonian involving three independent parameters expressed as a 7-point difference operator. This operator is obtained by combining the two independent difference equations satisfied by the bivariate Meixner polynomials. By construction, the wavefunctions of the model are given in terms of these two-variable polynomials. The energies of the system are given by the non-negative integers $N=0,1,2,\ldots$ and exhibit a $(N+1)$-fold degeneracy. Using the raising and lowering relations for the two-variable Meixner polynomials, the constants of motion of the system are constructed and are shown to close onto the $\mathfrak{su}(2)$ commutation relations. In the continuum limit, in which the lattice parameter tends to zero, the model contracts to the standard quantum harmonic oscillator, as required for a discrete oscillator model. The contraction process is illustrated at the wavefunction level using the two-variable Charlier polynomials in an intermediary step. The continuum limit is also displayed at the level of operators.

Here is the outline of the paper. In Section two, the essential properties of the two-variable Meixner polynomials are reviewed. In Section three, the Hamiltonian of the model is defined, the constants of motion are constructed, and the wavefunctions are illustrated. In Section four, the continuum limit of the model and wavefunctions is examined. We conclude with an outlook.

\section{The two-variable Meixner polynomials}
We now review the properties of the two-variable Meixner polynomials using the formalism and notation developed in \cite{2014_Genest&Miki&Vinet&Zhedanov_JPhysA_47_045207}. Let $\beta \geqslant 0$ be a positive real number and let $\Lambda\in O(2,1)$ be a $3\times 3$ pseudo-rotation matrix. This implies that $\Lambda$ satisfies
\begin{align*}
 \Lambda^{\top} \,\eta \, \Lambda =\eta,
\end{align*}
where $\eta=\mathrm{diag}(1,1,-1)$ and where $\Lambda^{\top}$ denotes the transposed matrix. In general, $\Lambda$ can be parametrized by three real numbers akin to the Euler angles. The two-variable Meixner polynomials, denoted by $M_{n_1,n_2}^{(\beta)}(x_1,x_2)$, are defined by the generating function
\begin{multline}
\label{GenFun}
 \left(1+\frac{\Lambda_{11}}{\Lambda_{13}} z_1+\frac{\Lambda_{12}}{\Lambda_{13}}z_2\right)^{x_1}\left(1+\frac{\Lambda_{21}}{\Lambda_{23}} z_1+\frac{\Lambda_{22}}{\Lambda_{23}}z_2\right)^{x_2}\left(1+\frac{\Lambda_{31}}{\Lambda_{33}}z_1+\frac{\Lambda_{32}}{\Lambda_{33}}z_2\right)^{-x_1-x_2-\beta}
 \\
 =\sum_{n_1=0}^{\infty}\;\sum_{n_2=0}^{\infty} \sqrt{\frac{(\beta)_{n_1+n_2}}{n_1!n_2!}}\; M_{n_1,n_2}^{(\beta)}(x_1,x_2)\;z_1^{n_1}z_2^{n_2}, \qquad
\end{multline}
where the $\Lambda_{ij}$ are the entries of the parameter matrix $\Lambda$ and where $(\beta)_{n}$ stands for the Pochhammer symbol defined as
\begin{align*}
 (\beta)_n=
 \begin{cases}
  1 & n=0
  \\
  \prod_{k=0}^{n-1}(\beta+k) & n=1,2,3,\ldots
 \end{cases}
\end{align*}
It can be seen from \eqref{GenFun} that $M_{n_1,n_2}^{(\beta)}(x_1,x_2)$ are polynomials of total degree $n_1+n_2$ in the variables $x_1$ and $x_2$. The functions $M_{n_1,n_2}^{(\beta)}(x_1,x_2)$ satisfy the orthogonality relation
\begin{align}
\label{Ortho}
 \sum_{x_1=0}^{\infty}\sum_{x_2=0}^{\infty} \omega(x_1,x_2)\;M_{n_1,n_2}^{(\beta)}(x_1,x_2)\,M_{n_1',n_2'}^{(\beta)}(x_1,x_2)=\delta_{n_1,n_1'}\delta_{n_2,n_2'},
\end{align}
where $\omega(x_1,x_2)$ is the negative trinomial distribution
\begin{align}
\label{Weight}
 \omega(x_1,x_2)=\frac{(\beta)_{x_1+x_2}}{x_1!\,x_2!}\;(1-c_1-c_2)^{\beta}\;c_1^{x_1}c_2^{x_2},
\end{align}
and where the parameters $c_1$, $c_2$ are given by
\begin{align*}
 c_1=\left(\frac{\Lambda_{13}}{\Lambda_{33}}\right)^2,\qquad c_2=\left(\frac{\Lambda_{23}}{\Lambda_{33}}\right)^2.
\end{align*}
The polynomials $M_{n_1,n_2}^{(\beta)}(x_1,x_2)$ have an explicit expression in terms of Aomoto--Gelfand hypergeometric series \cite{2012_Iliev_AdvApplMath_49_15}. This expression reads
\begin{multline}
\label{Explicit}
 M_{n_1,n_2}^{(\beta)}(x_1,x_2)=(-1)^{n_1+n_2}\sqrt{\frac{(\beta)_{n_1+n_2}}{n_1!n_2!}} \left(\frac{\Lambda_{31}}{\Lambda_{33}}\right)^{n_1}\left(\frac{\Lambda_{32}}{\Lambda_{33}}\right)^{n_2}
 \\
 \times \sum_{\mu,\nu,\rho,\sigma\geqslant 0} \frac{(-n_1)_{\mu+\nu}(-n_2)_{\rho+\sigma}(-x_1)_{\mu+\rho}(-x_2)_{\nu+\sigma}}{\mu!\nu!\rho!\sigma! \;(\beta)_{\mu+\nu+\rho+\sigma}}\;(1-u_{11})^{\mu}(1-u_{21})^{\nu}(1-u_{12})^{\rho}(1-u_{22})^{\sigma},
\end{multline}
where the parameters $u_{ij}$ are given by
\begin{align*}
 u_{11}=\frac{\Lambda_{11}\Lambda_{33}}{\Lambda_{13}\Lambda_{31}},\quad u_{12}=\frac{\Lambda_{12}\Lambda_{33}}{\Lambda_{13}\Lambda_{32}},\quad u_{21}=\frac{\Lambda_{21}\Lambda_{33}}{\Lambda_{23}\Lambda_{31}},\quad u_{22}=\frac{\Lambda_{22}\Lambda_{33}}{\Lambda_{23}\Lambda_{32}}.
\end{align*}

Let $T_{x_i}^{\pm} f(x_i)=f(x_i\pm 1)$ for $i=1,2$ be the discrete shift operators in the variables $x_1$ and $x_2$. Introduce the two intertwining operators $A_{+}^{(i)}$ defined as
\begin{align}
\label{Raising}
 A_{+}^{(i)}&=\frac{\Lambda_{1i}}{\Lambda_{13}}\;x_1 T_{x_1}^{-}+\frac{\Lambda_{2i}}{\Lambda_{23}}\;x_2 T_{x_2}^{-}-\frac{\Lambda_{3i}}{\Lambda_{33}}\;(x_1+x_2+\beta)\,\mathbb{I},\qquad i=1,2,
\end{align}
where $\mathbb{I}$ stands for the identity operator. On the bivariate polynomials $M_{n_1,n_2}^{(\beta)}(x_1,x_2)$, these operators have the actions
\begin{align}
\label{Raise-Action}
\begin{aligned}
 A_{+}^{(1)}\,M_{n_1,n_2}^{(\beta+1)}(x_1,x_2)&=\sqrt{\beta\;(n_1+1)}\,M_{n_1+1,n_2}^{(\beta)}(x_1,x_2),
 \\
 A_{+}^{(2)}\,M_{n_1,n_2}^{(\beta+1)}(x_1,x_2)&=\sqrt{\beta\; (n_2+1)}\,M_{n_1,n_2+1}^{(\beta)}(x_1,x_2).
 \end{aligned}
\end{align}
Introduce also the two intertwining operators $A_{-}^{(i)}$ defined in the following way:
\begin{align}
\label{Lowering}
 A_{-}^{(i)}=\Lambda_{1i}\Lambda_{13}\;T_{x_1}^{+}+\Lambda_{2i}\Lambda_{23}\;T_{x_2}^{+}-(\Lambda_{1i}\Lambda_{13}+\Lambda_{2i}\Lambda_{23})\,\mathbb{I},\qquad i=1,2.
\end{align}
These operators act as follows on the bivariate Meixner polynomials
\begin{align}
\label{Lower-Action}
\begin{aligned}
 A_{-}^{(1)}\,M_{n_1,n_2}^{(\beta)}(x_1,x_2)&=\sqrt{\frac{n_1}{\beta}}\,M_{n_1-1,n_2}^{(\beta+1)}(x_1,x_2),
 \\
 A_{-}^{(2)}\,M_{n_1,n_2}^{(\beta)}(x_1,x_2)&=\sqrt{\frac{n_2}{\beta}}\, M_{n_1,n_2-1}^{(\beta+1)}(x_1,x_2).
 \end{aligned}
\end{align}
The intertwining operators \eqref{Raising} and \eqref{Lowering} can be combined to produce the two commuting difference operators $Y_1$ and $Y_2$ that are diagonalized by the bivariate Meixner polynomials.  These operators are defined as
\begin{align}
\label{Y2}
 Y_i=A_{+}^{(i)}A_{-}^{(i)},\qquad i=1,2.
\end{align}
Explicitly, one finds
\begin{multline}
\label{Y}
 Y_i=\left(\frac{\Lambda_{1i}\Lambda_{2i}\Lambda_{23}}{\Lambda_{13}}\right)\,x_1 T_{x_1}^{-}T_{x_2}^{+}+\left(\frac{\Lambda_{1i}\Lambda_{2i}\Lambda_{13}}{\Lambda_{23}}\right)\,x_2 T_{x_1}^{+}T_{x_2}^{-}
 -\left(\frac{\Lambda_{1i}\Lambda_{3i}\Lambda_{33}}{\Lambda_{13}}\right)\,x_1 T_{x_1}^{-}
 \\
 -\left(\frac{\Lambda_{2i}\Lambda_{3i}\Lambda_{33}}{\Lambda_{23}}\right)\,x_2 T_{x_2}^{-}
 -\left(\frac{\Lambda_{1i}\Lambda_{3i}\Lambda_{13}}{\Lambda_{33}}\right)\,(x_1+x_2+\beta)\,T_{x_1}^{+}
 \\
 -\left(\frac{\Lambda_{2i}\Lambda_{3i}\Lambda_{23}}{\Lambda_{33}}\right)\,(x_1+x_2+\beta)\,T_{x_2}^{+}+\left[\Lambda_{1i}^2\;x_1+\Lambda_{2i}^2\;x_2+\Lambda_{3i}^2\;(x_1+x_2+\beta)\right]\,\mathbb{I}.
\end{multline}
The eigenvalue equations read
\begin{align}
\label{Eigenvalue-Equations}
 Y_1 \,M_{n_1,n_2}^{(\beta)}(x_1,x_2)=n_1 \,M_{n_1,n_2}^{(\beta)}(x_1,x_2),\qquad 
 Y_2 \, M_{n_1,n_2}^{(\beta)}(x_1,x_2)=n_2 \,M_{n_1,n_2}^{(\beta)}(x_1,x_2),
\end{align}
where $n_1$, $n_2$ are non-negative integers. Let us note that the operators $Y_1$ and $Y_2$ fully characterize the polynomials $M_{n_1,n_2}^{(\beta)}(x_1,x_2)$.
\section{A discrete and superintegrable Hamiltonian}
We now consider the Hamiltonian obtained by taking the sum of the operators $Y_1$ and $Y_2$ involved in the eigenvalue equations \eqref{Eigenvalue-Equations}. We hence define
\begin{align}
\label{Hamiltonian}
 \mathcal{H}=Y_1+Y_2.
\end{align}
In principle the Hamiltonian \eqref{Hamiltonian} involves four independent parameters including $\beta$, as any matrix $\Lambda\in O(2,1)$ depends on three independent parameters. However, it turns out that $\mathcal{H}$ essentially depends only on three parameters, including $\beta$. This can be seen explicitly as follows. Consider the following parametrization of $\Lambda$ in terms of the ``Euler angles'' $\psi$, $\xi$ and $\phi$:
\begin{align}
\label{Para}
 \Lambda(\psi,\xi,\phi)=
 \begin{pmatrix}
  \cosh\xi & 0 & \sinh\xi
  \\
  0 & 1 & 0
  \\
  \sinh\xi & 0 &\cosh\xi
 \end{pmatrix}
 \begin{pmatrix}
  1 & 0 & 0
  \\
  0 & \cosh\psi & \sinh\psi
  \\
  0 & \sinh\psi & \cos\psi
 \end{pmatrix}
 \begin{pmatrix}
  \cos  \phi & \sin \phi & 0
  \\
  -\sin \phi & \cos \phi & 0
  \\
  0 & 0 & 1
 \end{pmatrix}.
\end{align}
Upon using the expressions \eqref{Y}, it is seen that in the parametrization \eqref{Para} the Hamiltonian \eqref{Hamiltonian} has the expression
\begin{multline*}
 \mathcal{H}(\psi, \xi,\phi)=\sinh^2 \psi\;x_1 T_{x_1}^{-}T_{x_2}^{+}+\cosh^2\psi \sinh^2 \xi \;x_2 T_{x_1}^{+}T_{x_2}^{-}
 \\
-\cosh^2\psi \sinh^2\xi\;(x_1+x_2+\beta)\,T_{x_1}^{+}-\sinh^2\psi\;(x_1+x_2+\beta)\,T_{x_2}^{+}
\\
-\cosh^2\xi \cosh^2\psi\;\big[x_1T_{x_1}^{-}+x_2 T_{x_2}^{-}\big]+\Big[\cosh 2\xi \cosh^2 \psi \;x_1
\\
+(\cosh^2\psi+\sinh^2\xi+\cosh^2\xi \sinh^2 \psi)\;x_2+\beta\;(\sinh^2\xi+\cosh^2\xi \sinh^2\psi)\Big]\,\mathbb{I},
\end{multline*}
hence the parameter $\phi$ does not explicitly appear in $\mathcal{H}$. The fact that the Hamiltonian \eqref{Hamiltonian} can be presented in terms three independent parameters is a manifestation of its superintegrability.  Let $J_{X}$, $J_{Y}$ and $J_{Z}$ be the operators defined as follows.
\begin{align}
\label{Realization}
\begin{aligned}
 J_{X}&=\frac{1}{2}\left(A_{+}^{(1)}A_{-}^{(2)}+A_{+}^{(2)}A_{-}^{(1)}\right),
 \\
 J_{Y}&=\frac{1}{2i}\left(A_{+}^{(1)}A_{-}^{(2)}-A_{+}^{(2)}A_{-}^{(1)}\right),
 \\
 J_{Z}&=\frac{1}{2}\left(A_{+}^{(1)}A_{-}^{(1)}-A_{+}^{(2)}A_{-}^{(2)}\right).
 \end{aligned}
\end{align}
The operators $J_{X}$, $J_{Y}$ and $J_{Z}$ are constants of motion. Indeed, one can verify by a direct calculation that these operators commute with the Hamiltonian \eqref{Hamiltonian}
\begin{align*}
 [\mathcal{H},J_{X}]=0,\qquad [\mathcal{H},J_{Y}]=0,\qquad [\mathcal{H},J_{Z}]=0.
\end{align*}
The symmetry operators $J_{X}$, $J_{Y}$ and $J_{Z}$ satisfy the defining relations of the $\mathfrak{su}(2)$ algebra. One has
\begin{align}
\label{Sym-Alg}
 [J_{X},J_{Y}]=i J_{Z},\qquad [J_{Y},J_{Z}]=i J_{X},\qquad [J_{Z},J_{X}]=i J_{Y}.
\end{align}
In the realization \eqref{Realization}, the $\mathfrak{su}(2)$ Casimir operator is related to the Hamiltonian \eqref{Hamiltonian} through
\begin{align}
\label{Cas-Relation}
 J_{X}^2+J_{Y}^2+J_{Z}^2=\frac{1}{2} \mathcal{H}\left(\frac{\mathcal{H}}{2}+1\right).
\end{align}
The realization \eqref{Realization} of the $\mathfrak{su}(2)$ algebra \eqref{Sym-Alg} and the expression \eqref{Cas-Relation} of the Casimir operator in terms of the Hamiltonian is very close to the Schwinger realization of $\mathfrak{su}(2)$ that one finds when considering the standard two-dimensional quantum harmonic oscillator. The $SU(2)$ symmetry \eqref{Sym-Alg} of the Hamiltonian \eqref{Hamiltonian} explains why $\mathcal{H}$ depends on three parameters instead of four: the $\phi$ parameter in \eqref{Hamiltonian} has been ``rotated away'' from the Hamiltonian by the choice of parametrization \eqref{Para}.

By construction, the eigenfunctions of the Hamiltonian \eqref{Hamiltonian} are expressed in terms of the two-variable Meixner polynomials $M_{n_1,n_2}^{(\beta)}(x_1,x_2)$. These eigenfunctions $\Psi_{N,n}^{(\beta)}(x_1,x_2)$ are labeled by the two non-negative integers $N$ and $n$ and read
\begin{align*}
 \Psi_{N,n}^{(\beta)}(x_1,x_2)=M_{n,N-n}^{(\beta)}(x_1,x_2),
\end{align*}
where $n=0,1,\ldots, N$ and where $N=0,1,2,\ldots$. One has
\begin{align}
\label{Action-1}
 \mathcal{H}\,\Psi_{N,n}^{(\beta)}(x_1,x_2)=N\,\Psi_{N,n}^{(\beta)}(x_1,x_2),\qquad J_{Z}\,\Psi_{N,n}^{(\beta)}(x_1,x_2)=(n-N/2)\,\Psi_{N,n}^{(\beta)}(x_1,x_2).
\end{align}
Hence the eigenvalues of $\mathcal{H}$ are the non-negative integers $N=0,1,2,\ldots$ and are $(N+1)$-times degenerate. The states $\Psi_{N,n}(x_1,x_2)$ support $(N+1)$-dimensional irreducible representations of $\mathfrak{su}(2)$. Upon introducing the generators
\begin{align*}
 J_{\pm }=J_{X}\pm i J_{Y},
\end{align*}
it is seen from \eqref{Raise-Action} and \eqref{Lower-Action} that these operators have the actions
\begin{align}
\label{Action-2}
 J_{+}\,\Psi_{N,n}^{(\beta)}(x_1,x_2)&=\sqrt{(n+1)(N-n)}\,\Psi_{N,n+1}^{(\beta)}(x_1,x_2),
 \\
 J_{-}\,\Psi_{N,n}^{(\beta)}(x_1,x_2)&=\sqrt{n(N-n+1)}\,\Psi_{N,n-1}^{(\beta)}(x_1,x_2).
\end{align}
It thus follows that the two-variable Meixner polynomials $M_{n_1,n_2}^{(\beta)}(x_1,x_2)$ support $(K+1)$-dimensional unitary representations of $\mathfrak{su}(2)$ where $K=n_1+n_2$.

In view of \eqref{Ortho} wavefunctions $\Psi_{N,n}(x_1,x_2)$ are not normalized on the infinite grid $(x_1,x_2)\in \mathbb{N}\times \mathbb{N}$ with respect to the standard uniform measure of quantum mechanics. Properly normalized wavefunctions $\Upsilon_{N,n}^{(\beta)}(x_1,x_2)$ are obtained by taking
\begin{align}
\label{Wavefunctions}
 \Upsilon_{N,n}^{(\beta)}(x_1,x_2)=\sqrt{\omega(x_1,x_2)}\,M_{n,N-n}^{(\beta)}(x_1,x_2),
\end{align}
where $\omega(x_1,x_2)$ is given by \eqref{Weight}. One then has the orthogonality and completeness relations  \cite{2014_Genest&Miki&Vinet&Zhedanov_JPhysA_47_045207}
\begin{align*}
 \sum_{x_1=0}^{\infty}\sum_{x_2=0}^{\infty} \Upsilon_{N,n}^{(\beta)}(x_1,x_2)\Upsilon_{N',n'}^{(\beta)}(x_1,x_2)=\delta_{nn'}\delta_{NN'},
 \\
 \sum_{N=0}^{\infty}\sum_{n=0}^{N}\Upsilon_{N,n}^{(\beta)}(x_1,x_2)\Upsilon_{N,n}^{(\beta)}(x_1',x_2')=\delta_{x_1,x_1'}\delta_{x_2,x_2'}.
\end{align*}
The actions \eqref{Action-1} and \eqref{Action-2} on the non-normalized wavefunctions $\Psi_{N,n}^{(\beta)}(x_1,x_2)$ can be recovered on the normalized wavefunctions $\Upsilon_{N,n}^{(\beta)}(x_1,x_2)$ by applying the gauge transformation $\mathcal{O}\rightarrow \omega^{1/2}(x_1,x_2)\;\mathcal{O} \omega^{-1/2}(x_1,x_2)$, where $\mathcal{O}$ is either $\mathcal{H}$ or any one of the symmetries $J_{X}$, $J_{Y}$, $J_{Z}$.

Below are illustrated some of the wavefunctions amplitude $\rvert \Upsilon_{N,n}^{(\beta)}(x_1,x_2)\rvert $ for various values of the parameters $\xi$, $\psi$, $\theta$ and $\beta$. The model is defined on the $\mathbb{N}\times \mathbb{N}$ grid but only the grid $75\times 75$ is shown. It is seen that the particle is localized close to the origin. The energy level $N$ prescribes the number of ``nodes'' in the wavefunction amplitudes and the parameter $\beta$ is related to the spatial spreading of the amplitude.
\begin{center}
\scalebox{0.2}{\includegraphics{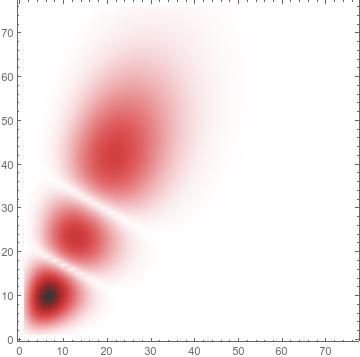}}\quad \scalebox{0.2}{\includegraphics{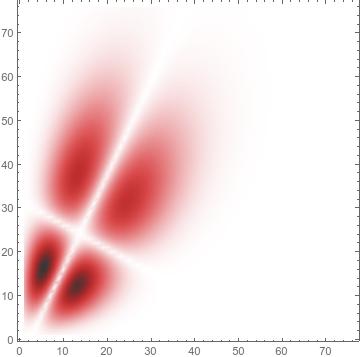}}\quad \scalebox{0.2}{\includegraphics{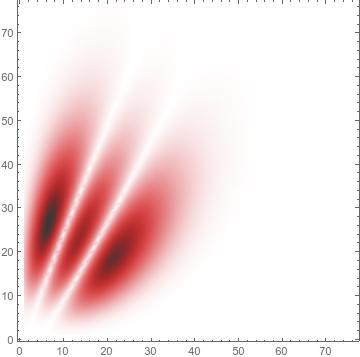}}
\\
 \scalebox{0.2}{\includegraphics{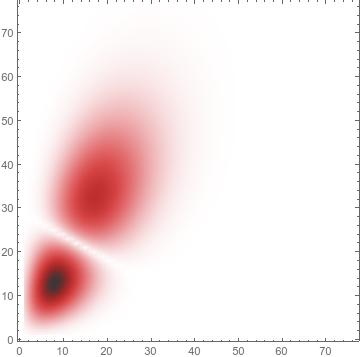}}\quad \scalebox{0.2}{\includegraphics{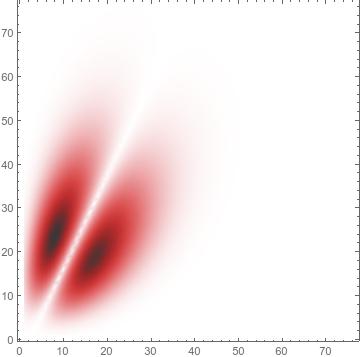}}
\\
 \scalebox{0.2}{\includegraphics{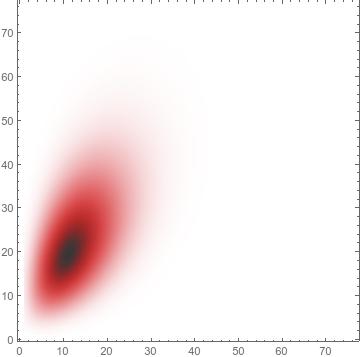}}
 \\
 Figure 1: Wavefunction amplitudes for $\xi=0.8$, $\psi=0.8$, $\phi=\pi/4$ and $\beta=15$. Bottom line: $|\Upsilon_{0,0}^{(\beta)}|$, middle line: $|\Upsilon_{1,0}^{(\beta)}|, |\Upsilon_{1,1}^{(\beta)}|$, top line: $|\Upsilon_{2,0}^{(\beta)}|,|\Upsilon_{2,1}^{(\beta)}|,|\Upsilon_{2,2}^{(\beta)}|$.
 \end{center}
 \begin{center}
 \scalebox{0.19}{\includegraphics{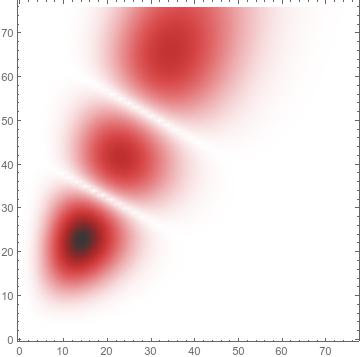}}\quad \scalebox{0.19}{\includegraphics{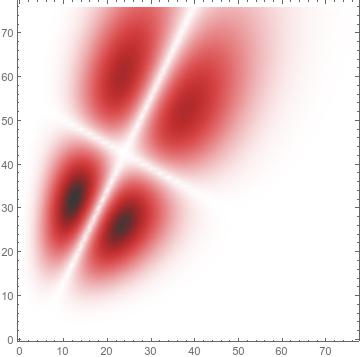}}\quad \scalebox{0.2}{\includegraphics{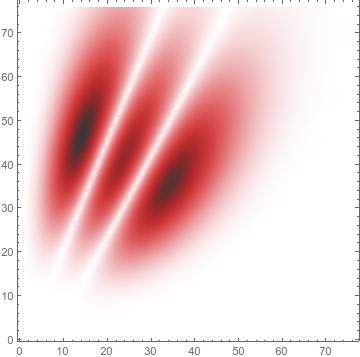}}
 \\
 \scalebox{0.19}{\includegraphics{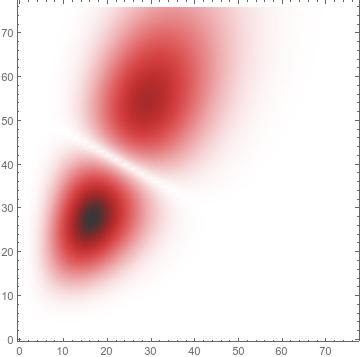}}\quad \scalebox{0.19}{\includegraphics{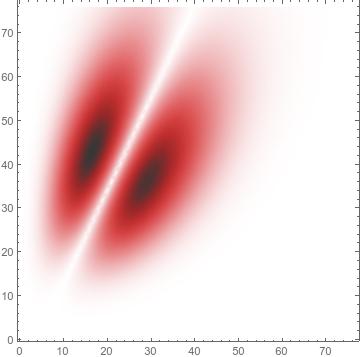}}
 \\
 \scalebox{0.19}{\includegraphics{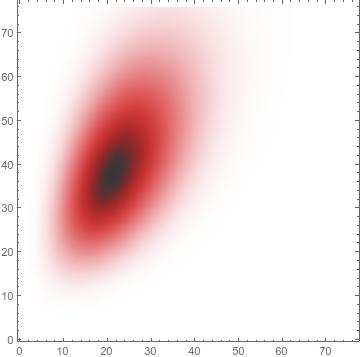}}
 \\
 Figure 2: Wavefunction amplitudes for $\xi=0.8$, $\psi=0.8$, $\phi=\pi/4$ and $\beta=28$. 
 Bottom line: $|\Upsilon_{0,0}^{(\beta)}|$, middle line: $|\Upsilon_{1,0}^{(\beta)}|, |\Upsilon_{1,1}^{(\beta)}|$, top line: $|\Upsilon_{2,0}^{(\beta)}|,|\Upsilon_{2,1}^{(\beta)}|,|\Upsilon_{2,2}^{(\beta)}|$.
 \end{center}
  \begin{center}
 \scalebox{0.19}{\includegraphics{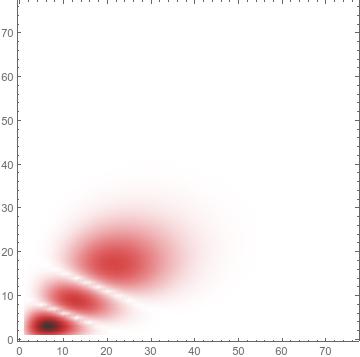}}\quad \scalebox{0.19}{\includegraphics{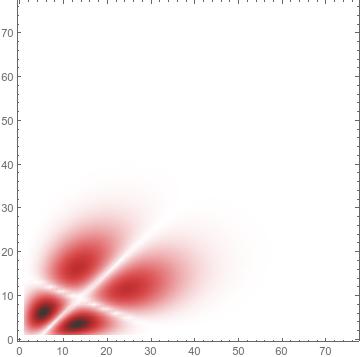}}\quad \scalebox{0.19}{\includegraphics{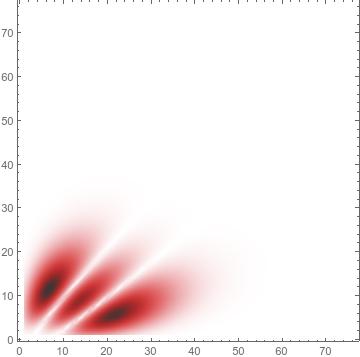}}
 \\
 \scalebox{0.19}{\includegraphics{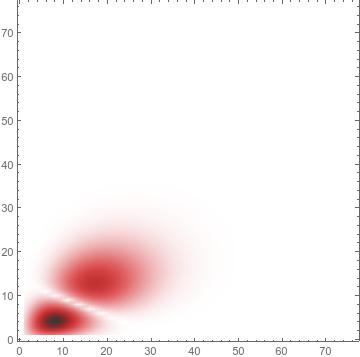}}\quad \scalebox{0.19}{\includegraphics{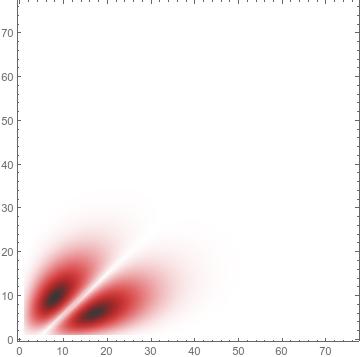}}
 \\
 \scalebox{0.19}{\includegraphics{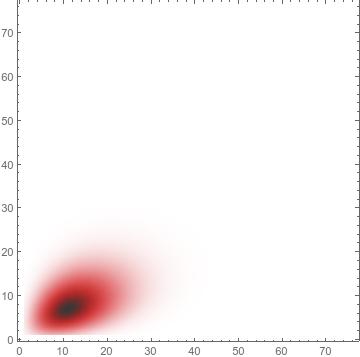}}
 \\
 Figure 3: Wavefunction amplitudes for $\xi=0.5$, $\psi=0.8$, $\phi=\pi/4$ and $\beta=15$. Bottom line: $|\Upsilon_{0,0}^{(\beta)}|$, middle line: $ |\Upsilon_{1,0}^{(\beta)}|, |\Upsilon_{1,1}^{(\beta)}|$, top line: $|\Upsilon_{2,0}^{(\beta)}|,|\Upsilon_{2,1}^{(\beta)}|,|\Upsilon_{2,2}^{(\beta)}|$.
 \end{center}
   \begin{center}
 \scalebox{.19}{\includegraphics{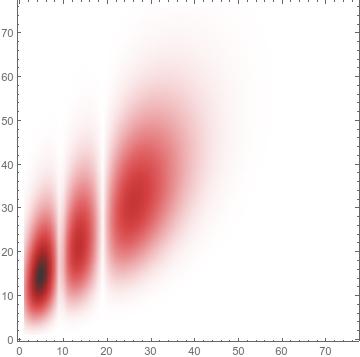}}\quad \scalebox{.19}{\includegraphics{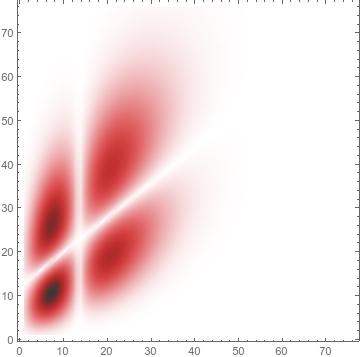}}\quad \scalebox{.19}{\includegraphics{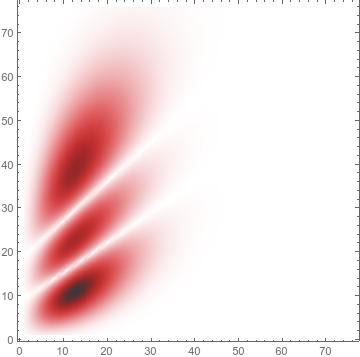}}
 \\
 \scalebox{.19}{\includegraphics{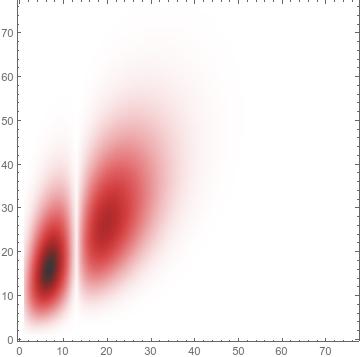}}\quad \scalebox{.19}{\includegraphics{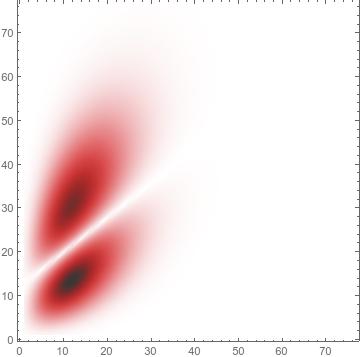}}
 \\
 \scalebox{.19}{\includegraphics{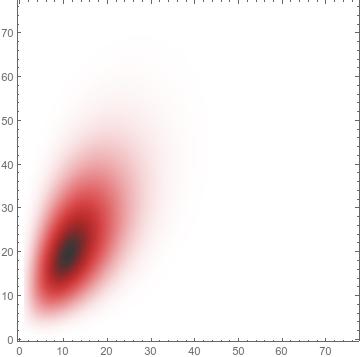}}
 \\
 Figure 4: Wavefunction amplitudes for $\xi=0.8$, $\psi=0.8$, $\phi=0$ and $\beta=15$. Bottom line: $|\Upsilon_{0,0}^{(\beta)}|$, middle line: $|\Upsilon_{1,0}^{(\beta)}|, |\Upsilon_{1,1}^{(\beta)}|$, top line: $|\Upsilon_{2,0}^{(\beta)}|,|\Upsilon_{2,1}^{(\beta)}|,|\Upsilon_{2,2}^{(\beta)}|$.
 \end{center}
\section{Continuum limit to the standard oscillator}
It will now be shown that in the continuum limit, the model governed by the Hamiltonian \eqref{Hamiltonian} tends to the standard isotropic quantum oscillator in two dimensions. The continuum limit from the two-variable Meixner polynomials to a product of two Hermite polynomials will be considered first. The explicit limit of the Hamiltonian \eqref{Hamiltonian} and of the constants of motion \eqref{Realization} will then be investigated.
\subsection{Continuum limit of the two-variable Meixner polynomials}
Consider the generating function \eqref{GenFun} of the two-variable Meixner polynomials \eqref{Explicit} in the parametrization \eqref{Para}. Upon writing
\begin{align}
\label{Lim1}
 \xi\rightarrow \frac{a}{\sqrt{\beta}},\qquad \psi\rightarrow \frac{b}{\sqrt{\beta}},\qquad z_1\rightarrow \frac{z_1}{\sqrt{\beta}},\qquad z_2\rightarrow \frac{z_2}{\sqrt{\beta}},
\end{align}
in left-hand side of \eqref{GenFun} and taking the limit as $\beta\rightarrow \infty$ using the standard result
\begin{align*}
 \lim_{k\rightarrow \infty}\left(1+\frac{x}{k}\right)^{k}=e^{x},
\end{align*}
one finds that
\begin{multline*}
 \lim_{\beta\rightarrow \infty}\Bigg[\left(1+\frac{\Lambda_{31}}{\Lambda_{33}}\frac{z_1}{\sqrt{\beta}}+\frac{\Lambda_{32}}{\Lambda_{33}}\frac{z_2}{\sqrt{\beta}}\right)^{-x_1-x_2-\beta}\\
 \times \left(1+\frac{\Lambda_{11}}{\Lambda_{13}} \frac{z_1}{\sqrt{\beta}}+\frac{\Lambda_{12}}{\Lambda_{13}}\frac{z_2}{\sqrt{\beta}}\right)^{x_1}\left(1+\frac{\Lambda_{21}}{\Lambda_{23}} \frac{z_1}{\sqrt{\beta}}+\frac{\Lambda_{22}}{\Lambda_{23}}\frac{z_2}{\sqrt{\beta}}\right)^{x_2}\Bigg]
 \\
 =\exp\left[-z_1(a\cos \phi-b\sin \phi)\right]
 \\
 \times \exp\left[-z_2(a\sin \phi+b \cos \phi)\right]\left(1+\frac{\cos \phi}{a} z_1+\frac{\sin \phi}{a} z_2\right)^{x_1}
 \left(1-\frac{\sin \phi}{b} z_1+\frac{\cos \phi}{b}z_2\right)^{x_2}.
\end{multline*}
Upon defining
\begin{align*}
 C_{n_1,n_2}(x_1,x_2)=\lim_{\beta\rightarrow \infty} M_{n_1,n_2}^{(\beta)}(x_1,x_2),
\end{align*}
and taking the limit as $\beta\rightarrow \infty$ with \eqref{Lim1} in the right-hand side of \eqref{GenFun}, one finds that
\begin{multline}
\label{GenFun-2}
 \exp\left[-z_1(a\cos \phi-b\sin \phi)\right]\exp\left[-z_2(a\sin \phi+b \cos \phi)\right] \left(1+\frac{\cos \phi}{a} z_1+\frac{\sin \phi}{a} z_2\right)^{x_1}\\
  \times\left(1-\frac{\sin \phi}{b} z_1+\frac{\cos \phi}{b}z_2\right)^{x_2}
 =\sum_{n_1=0}^{\infty}\sum_{n_2=0}^{\infty} C_{n_1,n_2}(x_1,x_2)\;\frac{z_1^{n_1}z_2^{n_2}}{\sqrt{n_1!n_2!}}.
\end{multline}
It is seen that the polynomials $C_{n_1,n_2}(x_1,x_2)$ correspond to the two-variable Charlier polynomials \cite{2014_Genest&Miki&Vinet&Zhedanov_JPhysA_47_215204}. If one uses the parametrization \eqref{Lim1} and takes the limit $\beta\rightarrow \infty$ in the weight function \eqref{Weight}, one finds the two-variable Poisson distribution
\begin{align}
\label{Poisson}
 \lim_{\beta\rightarrow \infty} \omega(x_1,x_2)=e^{-(a^2+b^2)}\frac{(a^2)^{x_1}(b^2)^{x_2}}{x_1!x_2!},
\end{align}
and the orthogonality relation \eqref{Ortho} becomes
\begin{align*}
 \sum_{x_1=0}^{\infty}\sum_{x_2=0}^{\infty}\left[e^{-(a^2+b^2)}\frac{(a^2)^{x_1}(b^2)^{x_2}}{x_1!x_2!}\right]\;C_{n_1,n_2}(x_1,x_2)C_{n_1',n_2'}(x_1,x_2)=\delta_{n_1,n_1'}\delta_{n_2,n_2'}.
\end{align*}
It is directly seen from \eqref{GenFun-2} and the standard generating function for the one-variable Charlier polynomials \cite{2010_Koekoek_&Lesky&Swarttouw} that when $\phi=0$, one has
\begin{align}
\label{Ortho-2}
 C_{n_1,n_2}(x_1,x_2)\Bigg\rvert_{\phi=0}=\frac{(-1)^{n_1+n_2}}{a^{n_1}b^{n_2}}C_{n_1}(x_1;a^2)C_{n_2}(x_2;b^2),
\end{align}
where $C_{n}(x;a)$ is the one-variable Charlier polynomials. Thus, using the standard limit from the one-variable Charlier polynomials to the one-variable Hermite polynomials, one can set
\begin{align}
\label{lim2}
 x_1= \sqrt{2} a \widetilde{x}_1+a^2,\qquad x_2=\sqrt{2} b\widetilde{x}_2+b^2,\qquad \phi=0,
\end{align}
in \eqref{GenFun-2} and take the limit as $a\rightarrow \infty$ and $b\rightarrow \infty $ to find
\begin{align*}
 \lim_{a\rightarrow \infty}\lim_{b\rightarrow \infty}e^{-az_1}\left(1+\frac{z_1}{a}\right)^{x_1}e^{-bz_2}\left(1+\frac{z_2}{b}\right)^{x_2}=e^{-\frac{z_1^2}{2}+\sqrt{2}\;\widetilde{x}_1\; z_1}e^{-\frac{z_2^2}{2}+\sqrt{2}\;\widetilde{x}_2\;z_2}.
\end{align*}
Upon comparing with the well-known generating function for the Hermite polynomials \cite{2010_Koekoek_&Lesky&Swarttouw}, one finds that
\begin{align*}
 \lim_{a\rightarrow \infty}\lim_{b\rightarrow \infty} C_{n_1,n_2}(x_1,x_2)\Bigg\rvert_{\phi=0}= \sqrt{2^{n_1+n_2}n_1!n_2!}\;H_{n_1}(\widetilde{x}_1)H_{n_2}(\widetilde{x}_2),
\end{align*}
where $H_{n}(x)$ are the standard Hermite polynomials. With the parametrization \eqref{lim2}, it is easily shown using Stirling's approximation that the bivariate Poisson distribution appearing in \eqref{Ortho-2} converges to the normal distribution
\begin{align*}
 \lim_{a\rightarrow \infty} e^{-a^2}\frac{(a^2)^{x_1}}{x_1!}=\frac{e^{-\widetilde{x}_1^2}}{\sqrt{\pi}}.
\end{align*}
In summary, the wavefunctions \eqref{Wavefunctions} of the discrete two-dimensional system governed by the Hamiltonian \eqref{Hamiltonian} tend to a separated product of two univariate Hermite polynomials in the combined limiting process \eqref{Lim1} and \eqref{lim2}. This motivates calling \eqref{Hamiltonian} a discrete oscillator. For other limits of bivariate orthogonal polynomials, see \cite{2013_Area&Godoy_JPhysA_46_035202}.\\

\noindent \emph{Remark}\\
It is not needed to take $\phi=0$ in the second limiting process involving the two-variable Charlier polynomials. If one keeps $\phi$ arbitrary and performs the change of variable \eqref{lim2} and takes the limit as $a\rightarrow \infty$ and $b\rightarrow \infty$, one simply finds a product of Hermite polynomials in the rotated coordinates $\widehat{x}_1=\cos \phi\,\widetilde{x}_1-\sin\phi\,\widetilde{x}_2$ and $\widehat{x}_2=\sin \phi\,\widetilde{x}_1+\cos \phi\,\widetilde{x}_2$.

\subsection{Continuum limit of the raising/lowering operators}
Let us now examine the combined limiting procedures of the preceding Subsection and its effect on the defining operators of the discrete oscillator model defined by \eqref{Hamiltonian}. We first consider the raising and lowering operators \eqref{Raising} and \eqref{Lowering} of the bivariate Meixner polynomials. Under the gauge transformation $\widetilde{A}_{\pm}^{(i)}=\omega^{1/2}(x_1,x_2)\;A_{\pm}^{(i)}\;\omega^{-1/2}(x_1,x_2)$, these operators have the expressions
\begin{align*}
 \widetilde{A}_{+}^{(i)}&=\Lambda_{1i}\,\sqrt{x_1}\;T_{x_1}^{-}+\Lambda_{2i}\,\sqrt{x_2}\;T_{x_2}^{-}-\Lambda_{3i}\,\sqrt{x_1+x_2+\beta}\;\mathbb{I},
 \\
 \widetilde{A}_{-}^{(i)}&=\Lambda_{1i}\,\sqrt{x_1+1}\;T_{x_1}^{+}+\Lambda_{2i}\,\sqrt{x_2+1}\;T_{x_2}^{+}-\Lambda_{3i}\,\sqrt{x_1+x_2+\beta}\;\mathbb{I},
\end{align*}
for $i=1,2$. On the wavefunctions \eqref{Wavefunctions}, one has
\begin{align*}
 \widetilde{A}_{+}^{(1)}\;\Upsilon_{N,n}^{(\beta+1)}(x_1,x_2)&=\sqrt{n+1}\;\Upsilon_{N+1,n+1}^{(\beta)}(x_1,x_2),
 \\
 \widetilde{A}_{+}^{(2)}\;\Upsilon_{N,n}^{(\beta+1)}(x_1,x_2)&=\sqrt{N-n+1}\;\Upsilon_{N+1,n}^{(\beta)}(x_1,x_2),
\end{align*}
and
\begin{align*}
 \widetilde{A}_{-}^{(1)}\;\Upsilon_{N,n}^{(\beta)}(x_1,x_2)&=\sqrt{n}\;\Upsilon_{n-1,N-1}^{(\beta+1)}(x_1,x_2),
 \\
 \widetilde{A}_{-}^{(2)}\;\Upsilon_{N,n}^{(\beta)}(x_1,x_2)&=\sqrt{N-n}\;\Upsilon_{n,N-1}^{(\beta+1)}(x_1,x_2).
\end{align*}
Upon taking the parametrization \eqref{Lim1} and taking the limit as $\beta\rightarrow \infty$, the raising operators become
\begin{align}
\label{lim3}
\begin{aligned}
 a_{+}^{(1)}&=\lim_{\beta\rightarrow \infty} \widetilde{A}_{+}^{(1)}=\cos \phi\;\sqrt{x_1}\;T_{x_1}^{-}-\sin\phi\;\sqrt{x_2}\;T_{x_2}^{-}-(a\cos\phi-b\sin \phi)\;\mathbb{I},
 \\
 a_{+}^{(2)}&=\lim_{\beta\rightarrow \infty}\widetilde{A}_{+}^{(2)}=\sin \phi\;\sqrt{x_1}\;T_{x_1}^{-}+\cos \phi \;\sqrt{x_2}\;T_{x_2}^{-}-(a\sin\phi+b\cos\phi)\;\mathbb{I},
 \end{aligned}
\end{align}
and the lowering operators become
\begin{align}
\label{lim4}
\begin{aligned}
 a_{-}^{(1)}&=\lim_{\beta\rightarrow \infty}\widetilde{A}_{-}^{(1)}=\cos\phi\;\sqrt{x_1+1}\;T_{x_1}^{+}-\sin\phi\;\sqrt{x_2+1}\;T_{x_2}^{+}-(a\cos\phi-b\sin \phi)\;\mathbb{I},
 \\
 a_{-}^{(2)}&=\lim_{\beta\rightarrow \infty}\widetilde{A}_{-}^{(1)}=\sin\phi\;\sqrt{x_1+1}\;T_{x_1}^{+}+\cos\phi\;\sqrt{x_2+1}\;T_{x_2}^{+}-(a\sin\phi+b\cos \phi)\;\mathbb{I}.
 \end{aligned}
\end{align}
A direct calculation shows that these operators satisfy the commutation relations
\begin{align*}
 [a_{-}^{(i)},a_{+}^{(j)}]=\delta_{ij},\qquad [a_{-}^{(i)},a_{-}^{(j)}]=0,\qquad [a_{+}^{(i)},a_{+}^{(j)}]=0.
\end{align*}
Upon setting $x_1= \sqrt{2} a \widetilde{x}_1+a^2$ and $x_2=\sqrt{2} b\widetilde{x}_2+b^2$ as in \eqref{lim2} and taking the limit as $a\rightarrow \infty$ and $b\rightarrow \infty$, it is easily seen that the operators \eqref{lim3} and \eqref{lim4} become the rotated creation/annihilation operators
\begin{align*}
 a_{+}^{(1)}\rightarrow \cos \phi\;a_1^{\dagger}-\sin \phi\;a_2^{\dagger},\quad a_{+}^{(2)}\rightarrow \sin \phi\;a_{1}^{\dagger}+\cos\phi\;a_{2}^{\dagger},
 \\
  a_{-}^{(1)}\rightarrow \cos \phi\;a_1-\sin \phi\;a_2,\quad a_{-}^{(2)}\rightarrow \sin \phi\;a_{1}+\cos\phi\;a_{2},
\end{align*}
where
\begin{align*}
 a_i=\frac{x_i+\partial_{x_i}}{\sqrt{2}},\qquad a_i^{\dagger}=\frac{x_i-\partial_{x_i}}{\sqrt{2}},
\end{align*}
are the usual creation/annihilation operators. It immediately follows that in the continuum limit described above in a two-step process, the gauge-transformed Hamiltonian \eqref{Hamiltonian} and constants of motion \eqref{Realization} obtained through $\omega(x_1,x_2)^{1/2}\mathcal{O}\omega^{-1/2}(x_1,x_2)$ tend to the standard two-dimensional oscillator Hamiltonian and the $\mathfrak{su}(2)$ generators in the Schwinger realization.
\section{Conclusion}
In this paper, we have introduced and described a discrete model of the oscillator in two-dimensions based on the bivariate Meixner polynomials. We have shown that this system is superintegrable and that it has the same symmetry algebra as its continuum limit, the standard isotropic oscillator in two dimensions. We have established that the two-variable Meixner polynomials form bases for irreducible representations of $\mathfrak{su}(2)$. We have detailed the limiting processes by which the two-variable Meixner polynomials tend to the bivariate Charlier polynomials and by which the latter tend to a product of standard Hermite polynomials.

In the present paper we have considered for simplicity the two-dimensional case. However, since the theory of multi-variable Meixner is now well established, it is clear that the model can be generalized to any dimensions to give a $d$-dimensional model of the harmonic oscillator with the same $\mathfrak{su}(d)$ symmetry as the standard quantum oscillator in $d$-dimensions. Another possible generalization would be to consider, instead of \eqref{Hamiltonian}, a discrete anisotropic oscillator with a Hamiltonian of the form $\mathcal{H}=Y_1+\alpha Y_2$. It is clear that for rational values of $\alpha$ this model would still be superintegrable, but would exhibit a higher order symmetry algebra. The case $\alpha=2$, which would correspond to a discrete $2:1$ anisotropic oscillator, is of particular interest in that regard \cite{1975_Boyer&Wolf_JMathPhys_16_2215}.
\ack
While this research was conducted, both JG and JL held an Undergraduate Student Research Award (USRA) from the Natural Sciences and Engineering Research Council of Canada (NSERC). VXG holds an Alexander--Graham--Bell PhD fellowship from NSERC. The research of LV is supported in part by NSERC.

\section*{References}

\end{document}